\documentclass[12pt]{article}
\pdfoutput=1
\usepackage{amsmath}
\def\be{\begin{equation}}
\def\ee{\end{equation}}
\def\arr{\begin{array}{rll}}
\def\ea{\end{array}}
\def\bea{\begin{eqnarray}}
\def\eea{\end{eqnarray}}

\def\N2{$N{=}2$}

\def\>{\rangle}
\def\<{\langle}
\def\+{\dagger}
\def\={\ =\ }

\begin{document}
\renewcommand{\thefootnote}{\fnsymbol{footnote}}
\begin{titlepage}
\setcounter{page}{0}
\vskip 1cm
\begin{center}
{\LARGE\bf  Integrability of $\mathcal{N}=1$ supersymmetric }\\
\vskip 0.4cm
{\LARGE\bf   Ruijsenaars--Schneider three--body system}\\
\vskip 1cm
$
\textrm{\Large Anton Galajinsky \ }
$
\vskip 0.7cm
{\it
Laboratory of Applied Mathematics and Theoretical Physics, \\
TUSUR, Lenin ave. 40, 634050 Tomsk, Russia} \\

\vskip 0.2cm
{e-mail: a.galajinsky@tusur.ru}
\vskip 0.5cm
\end{center}
\vskip 1cm
\begin{abstract} \noindent
An $\mathcal{N}=1$ 
supersymmetric extension of the Ruijsenaars-Schneider three--body model is constructed and its integrability is 
established. In particular, three functionally independent Grassmann--odd constants of the motion 
are given and their algebraic resolvability is proven.
The supersymmetric generalization is used to build a novel 
integrable isospin extension of the Ruijsenaars-Schneider three--body system.

\end{abstract}


\end{titlepage}

\renewcommand{\thefootnote}{\arabic{footnote}}
\setcounter{footnote}0

\noindent
{\bf 1. Introduction}\\

\noindent
Over the last few decades, supersymmetric extensions of integrable mechanics  
attracted a great deal of attention. On the one hand, such systems describe 
zero modes of superstring theories, which allow one to study complicated issues like the
kappa symmetry in a simpler setting \cite{GSW}. On the other hand, many--body models 
of that kind are believed to be relevant for a microscopic description of near horizon
black hole geometries \cite{CDKKTVP,GT}. 

An interrelation between supersymmetry and integrability 
received much less attention. It is widely
believed that a supersymmetric extension of an integrable system should automatically result
in a new integrable model. If this were the case, the supersymmetrization procedure would
provide an efficient means of generating novel integrable theories. However, 
because the number of fermionic degrees of freedom in general exceeds the number of
conserved supercharges, integrability of the resulting supersymmetric 
system is not a priori guaranteed. In particular, a recent analysis of an $\mathcal{N}=1$ 
supersymmetric extension of the Euler top \cite{AG1} indicates the lack of integrability in 
the fermionic sector. An alternative viewpoint on integrability of an $\mathcal{N}=2$ 
supersymmetric Euler top was presented in \cite{NHK}.

There is a subtlety regarding integrability in a fermionic sector of a supersymmetric
classical mechanics system. In general, it is not enough to construct 
functionally independent constants of the motion whose number 
is equal to the number of fermionic degrees of freedom.
Because fermionic integrals of motion are built from monomials in 
Grassmann--odd variables 
and there does not exist a division by a Grassmann--odd quantity \cite{BK}, 
in order to ensure integrability in the fermionic sector one has to find Grassmann--odd constants of the motion, 
which are {\it algebraically resolvable} with 
respect to the fermionic variables. A necessary condition for this is the presence 
of {\it linear} terms in each Grassmann--odd integral of motion. It is widely underestimated that 
fermionic constants of the motion, the leading term of which is quadratic in Grassmann--odd variables or higher, 
do not appear to be of practical utility at the classical level (see e.g. \cite{KLS} and 
references therein).

Aiming at a better understanding of an interplay between supersymmetry and 
integrability, in this work we systematically study integrability of an $\mathcal{N}=1$ 
supersymmetric extension of the Ruijsenaars-Schneider model \cite{RS} for the simplest case of 
three interacting particles. Although $\mathcal{N}=2$ Ruijsenaars-Schneider models are known 
in the literature \cite{BDM,AG,KL,KKL}, in this paper we choose to keep the number of 
fermions to a minimum. 

Another objective of this work is to emphasize that a nontrivial 
isospin extension of a bosonic many--body integrable model can be automatically achieved 
by properly reducing its $\mathcal{N}=1$ supersymmetric generalization.

The work is organized as follows. In the next section, we briefly 
review the Ruijsenaars-Schneider model and 
its integrals of motion. In Sect. 3, an $\mathcal{N}=1$ 
supersymmetric extension of the Ruijsenaars-Schneider three--body system is 
constructed and its integrability is established. Both the Lagrangian 
and Hamiltonian descriptions are presented. In particular, three functionally 
independent Grassmann--odd constants of the motion 
are given and their algebraic resolvability is proven.
The issue of constructing a Lax pair is discussed as well.
In Sect. 4, the $\mathcal{N}=1$ 
supersymmetry is used to build an integrable isospin extension of 
the original Ruijsenaars-Schneider three--body model. In the concluding Sect. 5, we summarize 
our results and discuss issues deserving of further study.  
Equations of motion for some subsidiary functions, which are extensively used in Sect. 3, are 
gathered in Appendix.

Throughout the paper summation over repeated indices is understood
unless otherwise stated.

\vspace{0.5cm}

\noindent
{\bf 2. The Ruijsenaars-Schneider models}\\

The Ruijsenaars-Schneider model is described by the equations of motion \cite{RS}
\be\label{RS}
\ddot{x}_i=\sum_{j\ne i} {\dot x}_i {\dot x}_j W(x_{ij}),
\ee
where $i,j=1,\dots,n$, $x_{ij}=x_i-x_j$, and $W(x)=\frac{2}{x}$, $\frac{2}{\sinh{x}}$, $2 \coth{x}$, $\frac{2}{\sin{x}}$, $2 \cot{x}$.
The hyperbolic variants seem to be of primary concern because the rational system is 
equivalent to a free particle \cite{ABHL}, while the trigonometric versions follow from 
the hyperbolic models by the substitution $x_i\to {\rm i} x_i$. For definiteness, in what follows 
we focus on $W(x)=\frac{2}{\sinh{x}}$. Throughout the text,
the condition $x_1>x_2>\dots >x_n$ is assumed to be valid. 

For $W(x)=\frac{2}{\sinh{x}}$, the dynamical equations (\ref{RS}) 
can be put into the Lax form ${\dot L}=[L,M]$ by introducing two matrices \cite{C} (no sum over repeated indices)
\be\label{LP}
L_{ij}=\delta_{ij} {\dot x}_i +(1-\delta_{ij}) \sqrt{{\dot x}_i {\dot x}_j} 
{\left(\cosh{\left(\frac{x_{ij}}{2}\right)}\right)}^{-1}, 
\quad M_{ij}=-\frac 12 (1-\delta_{ij}) \sqrt{{\dot x}_i {\dot x}_j} {\left(\sinh{\left(\frac{x_{ij}}{2}\right)}\right)}^{-1}.
\ee
The Lax formalism provides an efficient means of building constants of the motion  
in terms of $\mbox{Tr} L^k$, $k=1,\dots,n$. In particular, for the three--body case one 
finds the first integrals
\bea\label{fi}
&& 
I_1={\dot x}_1+{\dot x}_2+{\dot x}_3, 
\nonumber\\[2pt] 
&&
I_2={\dot x}_1 {\dot x}_2 \tanh^2{\left( \frac{x_{12}}{2}\right)}+{\dot x}_1 {\dot x}_3 \tanh^2{\left( \frac{x_{13}}{2}\right)}+{\dot x}_2 {\dot x}_3 \tanh^2{\left( \frac{x_{23}}{2}\right)}, \nonumber\\[2pt] 
&&
I_3={\dot x}_1 {\dot x}_2 {\dot x}_3 \tanh^2{\left( \frac{x_{12}}{2}\right)} 
\tanh^2{\left( \frac{x_{13}}{2}\right)} \tanh^2{\left( \frac{x_{23}}{2}\right)},
\eea
which link to $\mbox{Tr} L^k$ via the identities
\be\label{Lid}
\mbox{Tr} L-I_1=0, \qquad 
\mbox{Tr} L^2-{\left(\mbox{Tr} L\right)}^2+2 I_2=0,
\qquad
\mbox{Tr} L^3-{\left(\mbox{Tr} L \right)}^3
+3 I_1 I_2-3 I_3=0.
\ee
Because $\mbox{rank} \left(\frac{\partial I_a}{\partial \Gamma_\alpha} \right)=3$, 
where $\Gamma_\alpha=(x_1,x_2,x_3,{\dot x}_1,{\dot x}_2,{\dot x}_3)$, constants of the motion (\ref{fi}) are functionally independent.

For a greater number of particles, integrals of motion can be built iteratively. For example, 
in order to describe the four--body case, it suffices to introduce a new dynamical pair 
$(x_4,{\dot x}_4)$, 
extend (\ref{fi}) in a way compatible with the permutation invariance 
\bea\label{fi1}
&& 
I_1={\dot x}_1+{\dot x}_2+{\dot x}_3+{\dot x}_4, 
\nonumber\\[2pt]
&&
I_2={\dot x}_1 {\dot x}_2 \tanh^2{\left( \frac{x_{12}}{2}\right)}
+{\dot x}_1 {\dot x}_3 \tanh^2{\left( \frac{x_{13}}{2}\right)}
+{\dot x}_1 {\dot x}_4 \tanh^2{\left( \frac{x_{14}}{2}\right)}
\nonumber\\[2pt]
&&
\quad
+{\dot x}_2 {\dot x}_3 \tanh^2{\left( \frac{x_{23}}{2}\right)}
+{\dot x}_2 {\dot x}_4 \tanh^2{\left( \frac{x_{24}}{2}\right)}
+{\dot x}_3 {\dot x}_4 \tanh^2{\left( \frac{x_{34}}{2}\right)}, 
\nonumber\\[2pt]
&&
I_3=
{\dot x}_1 {\dot x}_2 {\dot x}_3 \tanh^2{\left( \frac{x_{12}}{2}\right)}
\tanh^2{\left( \frac{x_{13}}{2}\right)} 
\tanh^2{\left( \frac{x_{23}}{2}\right)}
\nonumber\\[2pt] 
&&
\qquad
+
{\dot x}_1 {\dot x}_2 {\dot x}_4 \tanh^2{\left( \frac{x_{12}}{2}\right)}
\tanh^2{\left( \frac{x_{14}}{2}\right)}
\tanh^2{\left( \frac{x_{24}}{2}\right)}
\nonumber\\[2pt] 
&&
\qquad
+{\dot x}_1 {\dot x}_3 {\dot x}_4 \tanh^2{\left( \frac{x_{13}}{2}\right)} 
\tanh^2{\left( \frac{x_{14}}{2}\right)}
\tanh^2{\left( \frac{x_{34}}{2}\right)}
\nonumber\\[2pt] 
&&
\qquad
+{\dot x}_2 {\dot x}_3 {\dot x}_4 \tanh^2{\left( \frac{x_{23}}{2}\right)}
\tanh^2{\left( \frac{x_{24}}{2}\right)}
\tanh^2{\left( \frac{x_{34}}{2}\right)},
\eea
and then add a four--body analogue of $I_3$ in (\ref{fi})
\bea\label{fi2}
&&	
I_4=
{\dot x}_1 {\dot x}_2 {\dot x}_3  {\dot x}_4 \tanh^2{\left( \frac{x_{12}}{2}\right)}
\tanh^2{\left( \frac{x_{13}}{2}\right)}
\tanh^2{\left( \frac{x_{14}}{2}\right)}
\tanh^2{\left( \frac{x_{23}}{2}\right)}
\nonumber\\[2pt] 
&&
\qquad 
\times
\tanh^2{\left( \frac{x_{24}}{2}\right)}
\tanh^2{\left( \frac{x_{34}}{2}\right)}.
\eea
Like in the preceding case, one can verify that  (\ref{fi1}), (\ref{fi2}) are functionally independent.
A five--body model is similarly built upon the four--body system and so on and so forth.

\vspace{0.5cm}

\noindent
{\bf 3. Integrability of $\mathcal{N}=1$ supersymmetric Ruijsenaars-Schneider three--body model}\\

\noindent
{\it 3.1 The Hamiltonian framework}\\

Perhaps the most straightforward way to construct an $\mathcal{N}=1$ supersymmetric extension
of the Ruijsenaars-Schneider model is to make recourse to the Hamiltonian 
formalism. It suffices to associate a real fermionic 
partner $\theta_i$ to each bosonic canonical pair $(x_i,p_i)$, $i=1,\dots,n$. 
The fermions are assumed to be self--conjugate\footnote{When passing to the Hamiltonian formalism, 
the standard fermionic kinetic term 
$\frac{\rm i}{2} \int dt \theta_i {\dot\theta}_i$ gives rise to the second class constraints
$p_{\theta i}-\frac{\rm i}{2} \theta_i=0$, where 
$p_{\theta i}=\frac{\partial \mathcal{L}}{\partial \theta_i}$ is the momentum 
canonically conjugate to $\theta_i$, $\mathcal{L}=\frac{\rm i}{2} \theta_i {\dot\theta}_i$ 
and the right derivative with respect to the Grassmann--odd variables is used.  Introducing the conventional Dirac bracket and eliminating $p_{\theta i}$ from the consideration by resolving the second class constraints,
one arrives at (\ref{DB}).}
\be\label{DB}
\{\theta_i,\theta_j \}=-{\rm i} \delta_{ij}.
\ee 
Because the original bosonic Hamiltonian is the sum of non--negative terms 
$H_B=\lambda_i \lambda_i$ (see Eqs. (\ref{sf})--(\ref{HBB}) below), an $\mathcal{N}=1$ 
supersymmetry charge can be chosen in the simplest form $Q=\lambda_i \theta_i$, which is 
linear in the fermions.
Via the Poisson bracket $\{Q,Q \}=-{\rm i} \mathcal{H}$, 
it gives rise to the superextended 
Hamiltonian $\mathcal{H}$.\footnote{In what follows, superextensions of the 
original bosonic quantities will be denoted by the same letters written in the 
calligraphic style.} The latter is the sum of $H_B$, which determines the 
dynamics of the original bosonic model, and extra
fermionic contributions describing boson--fermion couplings. The goal of this section is to establish integrability of an $\mathcal{N}=1$ 
supersymmetric Ruijsenaars-Schneider three--body model within the Hamiltonian framework.

Introducing momenta $p_i$ canonically conjugate to the coordinates $x_i$, $i=1,2,3$, 
the conventional Poisson bracket 
$\{x_i,p_j \}=\delta_{ij}$, and the Hamiltonian \cite{C}
\bea\label{H}
&&
H_B=e^{p_1} \coth{\left(\frac{x_{12}}{2}\right)}\coth{\left(\frac{x_{13}}{2}\right)}+e^{p_2} 
\coth{\left(\frac{x_{12}}{2}\right)} \coth{\left(\frac{x_{23}}{2}\right)}
\nonumber\\[2pt]
&&
\qquad 
+e^{p_3} \coth{\left(\frac{x_{13}}{2}\right)}\coth{\left(\frac{x_{23}}{2}\right)}=I_1,
\eea
one can bring the original equations of motion (\ref{RS}) to the Hamiltonian form.
Interestingly enough, the Hamiltonian (\ref{H}) links to $I_1$ in the previous section, 
which is linear in velocities. 
The counterpart of $I_2$ in (\ref{fi}) has the form
\bea\label{I2}
&&
I_2=e^{p_1+p_2} \coth{\left(\frac{x_{13}}{2}\right)} \coth{\left(\frac{x_{23}}{2}\right)}+e^{p_1+p_3} \coth{\left(\frac{x_{12}}{2}\right)} \coth{\left(\frac{x_{23}}{2}\right)}
\nonumber\\[2pt]
&&
\qquad \quad 
+e^{p_2+p_3} \coth{\left(\frac{x_{12}}{2}\right)} \coth{\left(\frac{x_{13}}{2}\right)},
\eea
while the analogue of $I_3$ reads
\be\label{I3}
I_3=e^{p_1+p_2+p_3}.
\ee
It is readily verified that $(I_1,I_2,I_3)$ 
are mutually commuting and functionally independent, which guarantees the Liouville integrability.

In order to construct an $\mathcal{N}=1$ supersymmetric extension, one first introduces
three subsidiary functions \cite{AG}
\bea\label{sf}
&&
\lambda_1=e^{\frac{p_1}{2}} \sqrt{ \coth{\left(\frac{x_{12}}{2}\right)}\coth{\left(\frac{x_{13}}{2}\right)}}, \qquad 
\lambda_2=e^{\frac{p_2}{2}}\sqrt{ \coth{\left(\frac{x_{12}}{2}\right)} \coth{\left(\frac{x_{23}}{2}\right)}}, 
\nonumber\\[2pt]
&&
\lambda_3=e^{\frac{p_3}{2} } \sqrt{\coth{\left(\frac{x_{13}}{2}\right)}\coth{\left(\frac{x_{23}}{2}\right)}},
\eea
which obey the quadratic algebra
(no summation over repeated indices and $i\ne j$)
\be\label{All}
\{\lambda_i,\lambda_j\}=\frac{\lambda_i \lambda_j}{2 \sinh{x_{ij}}},
\ee
and then represents the bosonic Hamiltonian (\ref{H}) in the quadratic form
\be\label{HBB}
H_B=\lambda_i \lambda_i.
\ee
For most of the calculations to follow, it proves convenient to trade $p_i$ for $\lambda_i$,
which slightly modifies the canonical bracket (no summation over repeated indices)
\be 
\{x_i,\lambda_j \}=\frac 12 \delta_{ij} \lambda_j.
\ee
The Hamiltonian 
equations of motion for $x_i$ and $\lambda_i$ are presented in Appendix. 

Then one builds the supersymmetry generator 
\be\label{SC}
Q_1=\lambda_i \theta_i,
\ee
and determines the superextended Hamiltonian $\mathcal{H}$ by computing the Poisson bracket
\be\label{SH}
\{Q_1,Q_1 \}=-{\rm i} \mathcal{H}, \qquad
\mathcal{H}=\lambda_i \lambda_i+
\frac{{\rm i}}{2} \sum_{i\ne j}
\frac{\lambda_i \lambda_j \theta_i \theta_j}{\sinh{x_{ij}}}=\mathcal{I}_1.
\ee
The latter governs the dynamics of the resulting $\mathcal{N}=1$ supersymmetric system. 

Supersymmetry transformations 
$\delta A={\rm i} \{A,Q_1 \} \epsilon$, where $A$ is an arbitrary function defined on the 
superextended phase space and $\epsilon$ is an infinitesimal Grassmann--odd constant parameter, read (no summation over repeated indices in the second formula below)
\be
\delta_\epsilon \theta_i=\lambda_i \epsilon, \qquad 
\delta_\epsilon x_i=\frac{{\rm i}}{2} \lambda_i \theta_i \epsilon, \qquad
\delta_\epsilon \lambda_i=\frac{{\rm i}}{2} \sum_{j\ne i} \frac{\lambda_i \lambda_j \theta_j \epsilon}{\sinh{x_{ij}}}.
\ee
Together with the temporal translation $\delta_a A=\{A,\mathcal{H} \} a$, where $\mathcal{H}$ is the Hamiltonian in (\ref{SH}) 
and $a$ is an infinitesimal Grassmann--even constant parameter, they form an on--shell closed algebra $[\delta_{\epsilon_1},\delta_{\epsilon_2}]=\delta_a$, 
with $a={\rm i} \epsilon_1 \epsilon_2$.

Similarly to the original bosonic model, the supersymmetric extension (\ref{SH}) is invariant under spatial
translations $x'_i=x_i+b$, where $b$ is a constant parameter. In particular, 
$I_3$ 
in (\ref{I3}) maintains its form after the supersymmetrization
\be\label{II3}
\mathcal{I}_3=e^{p_1+p_2+p_3},
\ee
which links to the conservation of the total momentum in the bosonic sector,
whereas $I_2$ in (\ref{I2}) acquires the fermionic contributions
\bea\label{II2}
&&
\mathcal{I}_2={\left(\lambda_1 \lambda_2 \tanh{ \left(\frac{x_{12}}{2}\right)} \right)}^2
+{\left(\lambda_1 \lambda_3 \tanh{ \left(\frac{x_{13}}{2}\right)} \right)}^2
+{\left( \lambda_2 \lambda_3 \tanh{ \left(\frac{x_{23}}{2}\right)} \right)}^2
\nonumber\\[2pt]
&&
\qquad +\frac{{\rm i}}{\sinh{x_{12}}} \lambda_1 \lambda_2 \lambda_3^2 \tanh{\left(\frac{x_{13}}{2} \right)}
\tanh{\left(\frac{x_{23}}{2} \right)} \theta_1 \theta_2
\nonumber\\[2pt]
&&
\qquad 
-\frac{{\rm i}}{\sinh{x_{13}}}  \lambda_1 \lambda_3 \lambda_2^2 \tanh{\left(\frac{x_{12}}{2} \right)}
\tanh{\left(\frac{x_{23}}{2} \right)} \theta_1 \theta_3
\nonumber\\[2pt]
&&
\qquad +\frac{{\rm i}}{\sinh{x_{23}}} \lambda_2 \lambda_3 \lambda_1^2 \tanh{\left(\frac{x_{12}}{2} \right)}
\tanh{\left(\frac{x_{13}}{2} \right)} \theta_2 \theta_3,
\eea
which are needed in order to guarantee its conservation over time and the superinvariance
\be\label{supp}
\{\mathcal{I}_2,\mathcal{H}\}=0, \qquad \{\mathcal{I}_2, Q_1 \}=0. 
\ee
Note that the simplest way to determine the fermionic terms in (\ref{II2}) is to solve the second 
equation in (\ref{supp}).

Remarkably enough, like the superextended Hamiltonian $\mathcal{H}=\mathcal{I}_1$ is produced 
by the bracket of two supercharges $\{Q_1,Q_1 \}=-{\rm i} \mathcal{H}$,
$\mathcal{I}_2$ in (\ref{II2}) also admits the superpartner
\bea\label{QQ2}
&&
Q_2=\lambda_2 \lambda_3 \tanh{\left(\frac{x_{23}}{2} \right)} \theta_1
-\lambda_1 \lambda_3 \tanh{\left(\frac{x_{13}}{2} \right)} \theta_2
+\lambda_1 \lambda_2 \tanh{\left(\frac{x_{12}}{2} \right)} \theta_3,
\eea
which obeys the relations
\bea
&&
\{Q_2,Q_2\}=-{\rm i} \mathcal{I}_2, \qquad \{Q_2,\mathcal{H} \}=0, \qquad \{Q_1,Q_2 \}=-{\rm i} \sqrt{\mathcal{I}_3}.
\eea
In verifying the structure relations above, the following Poisson brackets
\bea
&&
\left\{ \lambda_1,\lambda_2 \tanh{\left(\frac{x_{12}}{2}\right)} \right\}=0, \qquad 
\left\{\lambda_1,\lambda_3 \tanh{\left(\frac{x_{13}}{2}\right)} \right\}=0, 
\qquad \left\{\lambda_2,\lambda_1 \tanh{\left(\frac{x_{12}}{2}\right)} \right\}=0, 
\nonumber\\[2pt]
&&
\left\{\lambda_2,\lambda_3 \tanh{\left(\frac{x_{23}}{2}\right)} \right\}=0,
\qquad 
\left\{\lambda_3,\lambda_1 \tanh{\left(\frac{x_{13}}{2}\right)} \right\}=0,
\qquad
\left\{\lambda_3,\lambda_2 \tanh{\left(\frac{x_{23}}{2}\right)} \right\}=0,
\nonumber
\eea
proved useful.

Note that a priori the existence of $Q_2$ is not guaranteed. As a matter of fact, we considered a 
generic ansatz $Q_2=A_1 (x,\lambda) \theta_1+A_2 (x,\lambda) \theta_2+A_3 (x,\lambda) \theta_3
+A_4 (x,\lambda) \theta_1 \theta_2 \theta_3$, reduced the condition $\{Q_2,\mathcal{H} \}=0$ to 
a set of partial differential equations for the coefficients $(A_1,A_2,A_3,A_4)$, and then 
succeeded in finding a particular solution (\ref{QQ2}).

In order to ensure complete integrability of the model in the fermionic sector, 
a third Grassmann--odd constant of the motion is needed, the leading term of 
which is
linear in the fermionic variables. It turns out that an efficient way to reveal 
such a constant is to make recourse to higher order fermionic invariants and compute their
Poisson brackets with 
$Q_1$ and $Q_2$.

Analyzing equations of motion in the Grassmann--odd sector 
(see Eq. (\ref{SUSYeq2}) below) and taking into account the equalities $\theta_1^2=\theta_2^2=\theta_3^2=0$, 
one concludes that the cubic combination
\be\label{cubic}
\Omega=\theta_1 \theta_2 \theta_3
\ee
is conserved over time: $\{\Omega,\mathcal{H} \}=0$. Bracketing $\Omega$ with $Q_1$, $Q_2$, 
one obtains two constants of the motion, which are quadratic in the fermions
\bea
&&
\Lambda_1={\rm i} \lambda_3 \theta_1 \theta_2-{\rm i} \lambda_2  \theta_1 \theta_3+
{\rm i} \lambda_1  \theta_2 \theta_3,
\nonumber\\[2pt]
&&
\Lambda_2=
{\rm i} \lambda_1 \lambda_2 \tanh{\left(\frac{x_{12}}{2} \right)}  \theta_1 \theta_2
+{\rm i} \lambda_1 \lambda_3 \tanh{\left(\frac{x_{13}}{2} \right)}  \theta_1 \theta_3
+{\rm i} \lambda_2 \lambda_3 \tanh{\left(\frac{x_{23}}{2} \right)}  \theta_2 \theta_3,
\nonumber\\[2pt]
&&
\{Q_1,\Omega\}=-\Lambda_1, \qquad \{Q_2,\Omega\}=-\Lambda_2.
\eea
Finally, evaluating Poisson brackets among $\Lambda_{1,2}$ and $Q_{1,2}$, 
one reveals the desired third Grassmann--odd invariant, which involves terms linear in 
the fermionic variables
\bea
&&
Q_3=\lambda_1 \left(\lambda_2^2 \tanh{\left(\frac{x_{12}}{2} \right)} 
+\lambda_3^2 \tanh{\left(\frac{x_{13}}{2} \right)} \right) \theta_1
-\lambda_2 \left(\lambda_1^2 \tanh{\left(\frac{x_{12}}{2} \right)} 
-\lambda_3^2 \tanh{\left(\frac{x_{23}}{2} \right)} \right) \theta_2
\nonumber\\[2pt]
&&
\quad 
-\lambda_3 \left(\lambda_1^2 \tanh{\left(\frac{x_{13}}{2} \right)} 
+\lambda_2^2 \tanh{\left(\frac{x_{23}}{2} \right)} \right) \theta_3
\nonumber\\[2pt]
&&
\quad 
-\frac{{\rm i}}{2} \lambda_1 \lambda_2 \lambda_3 
\left(\tanh{\left(\frac{x_{23}}{2} \right)} 
\left(\frac{1}{\sinh{x_{12}}}+\frac{1}{\sinh{x_{13}}} \right)
+\tanh{\left(\frac{x_{13}}{2} \right)} \left(\frac{1}{\sinh{x_{12}}}-\frac{1}{\sinh{x_{23}}} \right)
\right.
\nonumber\\[2pt]
&&
\quad 
\left.
-
\tanh{\left(\frac{x_{12}}{2} \right)} \left(\frac{1}{\sinh{x_{13}}}+\frac{1}{\sinh{x_{23}}} \right)
 \right) \theta_1 \theta_2 \theta_3.
\eea
It satisfies the structure relations
\bea
&&
\{Q_1,\Lambda_1 \}=0, \qquad \{Q_1,\Lambda_2 \}=-Q_3, \qquad
\{Q_2,\Lambda_1 \}=Q_3, \qquad \{Q_2,\Lambda_2 \}=0.
\eea
Poisson brackets of $\Lambda_{1,2}$ with $Q_3$ do not result in new constants of the motion
\be
\{Q_3,\Lambda_1 \} =Q_1 \sqrt{\mathcal{I}_3}-Q_2 \mathcal{H}, \qquad 
\{Q_3,\Lambda_2 \} =Q_1 \mathcal{I}_2-Q_2 \sqrt{\mathcal{I}_3},
\ee
which is consistent with the fact that $Q_{1,2,3}$ suffice to provide 
the general solution to
the fermionic equations of motion (see Eq. (\ref{SUSYeq2}) below).

It turns out that the problem of algebraic resolvability of the fermionic constants of the motion $(Q_1,Q_2,Q_3)$ 
simplifies within the Lagrangian framework, to which we devote the next section.

\vspace{0.5cm}

\noindent
{\it 3.2 The Lagrangian picture}\\

Taking into account the Hamiltonian equations for $x_i$ and $\lambda_i$, 
which are presented in Appendix,
differentiating ${\dot x}_i$ with respect to the temporal variable $t$, 
and expressing $\lambda_i$ in terms of $(x_i,{\dot x}_i,\theta_i)$, one gets the 
supersymmetric extension 
of the original bosonic equations of motion (\ref{RS})
\bea\label{SUSYeq1}
&&
{\ddot x}_1=\frac{2 {\dot x}_1 {\dot x}_2 }{\sinh{x_{12}}} + \frac{2 {\dot x}_1 {\dot x}_3 }{\sinh{x_{13}}} +
\frac{{\rm i} \sqrt{{\dot x}_1 {\dot x}_2 } }{4 \cosh{\frac{x_{12}}{2}}} 
\left(\frac{{\dot x}_1 +{\dot x}_2 }{\cosh{\frac{x_{12}}{2}}}+\frac{{\dot x}_3}{\cosh{\frac{x_{13}}{2}}\cosh{\frac{x_{23}}{2}}} \right) \theta_1 \theta_2
\nonumber\\[2pt]
&&
\qquad 
+
\frac{{\rm i} \sqrt{{\dot x}_1 {\dot x}_3 } }{4 \cosh{\frac{x_{13}}{2}}} 
\left(\frac{{\dot x}_1 +{\dot x}_3 }{\cosh{\frac{x_{13}}{2}}}
+\frac{{\dot x}_2}{\cosh{\frac{x_{12}}{2}}\cosh{\frac{x_{23}}{2}}} \right) \theta_1 \theta_3,
\nonumber\\[2pt]
&&
{\ddot x}_2=-\frac{2 {\dot x}_1 {\dot x}_2 }{\sinh{x_{12}}} + \frac{2 {\dot x}_2 {\dot x}_3 }{\sinh{x_{23}}} -
\frac{{\rm i} \sqrt{{\dot x}_1 {\dot x}_2 } }{4 \cosh{\frac{x_{12}}{2}}} 
\left(\frac{{\dot x}_1 +{\dot x}_2 }{\cosh{\frac{x_{12}}{2}}}
+\frac{{\dot x}_3}{\cosh{\frac{x_{13}}{2}}\cosh{\frac{x_{23}}{2}}} \right) \theta_1 \theta_2
\nonumber\\[2pt]
&&
\qquad 
+
\frac{{\rm i} \sqrt{{\dot x}_2 {\dot x}_3 } }{4 \cosh{\frac{x_{23}}{2}}} 
\left(\frac{{\dot x}_2 +{\dot x}_3 }{\cosh{\frac{x_{23}}{2}}}
+\frac{{\dot x}_1}{\cosh{\frac{x_{12}}{2}}\cosh{\frac{x_{13}}{2}}} \right) \theta_2 \theta_3,
\nonumber
\eea
\bea
&&
{\ddot x}_3=-\frac{2 {\dot x}_1 {\dot x}_3 }{\sinh{x_{13}}} - \frac{2 {\dot x}_2 {\dot x}_3 }{\sinh{x_{23}}} -
\frac{{\rm i} \sqrt{{\dot x}_1 {\dot x}_3 } }{4 \cosh{\frac{x_{13}}{2}}} 
\left(\frac{{\dot x}_1 +{\dot x}_3 }{\cosh{\frac{x_{13}}{2}}}
+\frac{{\dot x}_2}{\cosh{\frac{x_{12}}{2}}\cosh{\frac{x_{23}}{2}}} \right) \theta_1 \theta_3
\nonumber\\[2pt]
&&
\qquad 
-
\frac{{\rm i} \sqrt{{\dot x}_2 {\dot x}_3 } }{4 \cosh{\frac{x_{23}}{2}}} 
\left(\frac{{\dot x}_2 +{\dot x}_3 }{\cosh{\frac{x_{23}}{2}}}
+\frac{{\dot x}_1}{\cosh{\frac{x_{12}}{2}}\cosh{\frac{x_{13}}{2}}} \right) \theta_2 \theta_3.
\eea
Dynamics in the fermionic sector is governed by
\bea\label{SUSYeq2}
&&
{\dot\theta}_1=\frac{\sqrt{{\dot x}_1 {\dot x}_2 }}{\sinh{x_{12}}} \theta_2
+\frac{\sqrt{{\dot x}_1 {\dot x}_3 }}{\sinh{x_{13}}} \theta_3, \quad 
{\dot\theta}_2=-\frac{\sqrt{{\dot x}_1 {\dot x}_2 }}{\sinh{x_{12}}} \theta_1
+\frac{\sqrt{{\dot x}_2 {\dot x}_3 }}{\sinh{x_{23}}} \theta_3,
\nonumber\\[2pt]
&&
{\dot\theta}_3=-\frac{\sqrt{{\dot x}_1 {\dot x}_3 }}{\sinh{x_{13}}} \theta_1
-\frac{\sqrt{{\dot x}_2 {\dot x}_3 }}{\sinh{x_{23}}} \theta_2.
\eea
Note that, as a consequence of the canonical equations for $x_i$ (see Appendix), the velocities ${\dot x}_i$ 
are non--negative, which means that their square roots are well defined. 

The Lagrangian form of the bosonic first integrals is found by 
linking $\lambda_i$ to $(x_i,{\dot x}_i,\theta_i)$ 
(see Appendix)
\bea\label{LaFI}
&&
\mathcal{I}_1={\dot x}_1+{\dot x}_2+{\dot x}_3, 
\nonumber\\[2pt]
&&
\mathcal{I}_2={\dot x}_1 {\dot x}_2 \tanh^2{\left( \frac{x_{12}}{2}\right)}
+{\dot x}_1 {\dot x}_3 \tanh^2{\left( \frac{x_{13}}{2}\right)}
+{\dot x}_2 {\dot x}_3 \tanh^2{\left( \frac{x_{23}}{2}\right)}
\nonumber\\[2pt]
&&
\qquad \quad
-\frac{{\rm i}}{4} \sqrt{{\dot x}_1 {\dot x}_2 } \tanh{\left(\frac{x_{12}}{2} \right)} 
\left(\frac{{\dot x}_1 +{\dot x}_2}{\cosh^2 {\left(\frac{x_{12}}{2} \right)}}
+\frac{{\dot x}_3}{\cosh^2 {\left(\frac{x_{13}}{2} \right)} \cosh^2 {\left(\frac{x_{23}}{2} \right)}}\right) 
\theta_1 \theta_2
\nonumber\\[2pt]
&&
\qquad \quad
-\frac{{\rm i}}{4} \sqrt{{\dot x}_1 {\dot x}_3 } \tanh{\left(\frac{x_{13}}{2} \right)} 
\left(\frac{{\dot x}_1 +{\dot x}_3}{\cosh^2 {\left(\frac{x_{13}}{2} \right)}}
+\frac{{\dot x}_2}{\cosh^2 {\left(\frac{x_{12}}{2} \right)} \cosh^2 {\left(\frac{x_{23}}{2} \right)}}\right) 
\theta_1 \theta_3
\nonumber\\[2pt]
&&
\qquad \quad
-\frac{{\rm i} }{4} \sqrt{{\dot x}_2 {\dot x}_3 } \tanh{\left(\frac{x_{23}}{2} \right)} 
\left(\frac{{\dot x}_2 +{\dot x}_3}{\cosh^2 {\left(\frac{x_{23}}{2} \right)}}
+\frac{{\dot x}_1}{\cosh^2 {\left(\frac{x_{12}}{2} \right)} \cosh^2 {\left(\frac{x_{13}}{2} \right)}}\right) 
\theta_2 \theta_3,
\nonumber\\[2pt]
&&
\mathcal{I}_3=\tanh^2{\left( \frac{x_{12}}{2}\right)} \tanh^2{\left( \frac{x_{13}}{2}\right)} 
\tanh^2{\left( \frac{x_{23}}{2}\right)} \left( {\dot x}_1 {\dot x}_2 {\dot x}_3 
-\frac{{\rm i} \sqrt{{\dot x}_1 {\dot x}_2}}{2 \sinh{x_{12}}} \left({\dot x}_1+ {\dot x}_2 \right){\dot x}_3 
 \theta_1 \theta_2
\right.
\nonumber\\[2pt]
&&
\qquad \quad
\left.
-\frac{{\rm i} \sqrt{{\dot x}_1 {\dot x}_3} }{2 \sinh{x_{13}}} \left({\dot x}_1+ {\dot x}_3 \right){\dot x}_2 
\theta_1 \theta_3
-\frac{{\rm i} \sqrt{{\dot x}_2 {\dot x}_3} }{2 \sinh{x_{23}}} \left({\dot x}_2+ {\dot x}_3 \right){\dot x}_1 
 \theta_2 \theta_3
\right),
\eea
while the fermionic constants of the motion acquire the form
\bea\label{FIOM}
&&
Q_1=\sqrt{{\dot x}_1} \theta_1+\sqrt{{\dot x}_2} \theta_2+\sqrt{{\dot x}_3} \theta_3,
\nonumber\\[2pt]
&&
Q_2=\sqrt{{\dot x}_2 {\dot x}_3} \tanh{\left(\frac{x_{23}}{2}\right)} \theta_1
-\sqrt{{\dot x}_1 {\dot x}_3} \tanh{\left(\frac{x_{13}}{2}\right)} \theta_2
+\sqrt{{\dot x}_1 {\dot x}_2} \tanh{\left(\frac{x_{12}}{2}\right)} \theta_3
\nonumber\\[2pt]
&&
\qquad \quad
-\frac{{\rm i}}{8} \left(\frac{{\dot x}_1+ {\dot x}_2}{\cosh^2 {\left(\frac{x_{12}}{2} \right)}} +
\frac{{\dot x}_1+ {\dot x}_3}{\cosh^2 {\left(\frac{x_{13}}{2} \right)}}
+\frac{{\dot x}_2+ {\dot x}_3}{\cosh^2 {\left(\frac{x_{23}}{2} \right)}}
\right) \theta_1 \theta_2 \theta_3,
\nonumber
\eea
\bea
&&
Q_3=\sqrt{\dot{x}_1} \left(\dot{x}_2 \tanh{\left(\frac{x_{12}}{2} \right)} 
+\dot{x}_3 \tanh{\left(\frac{x_{13}}{2} \right)} \right) \theta_1
\nonumber\\[2pt]
&&
\qquad \quad 
-\sqrt{\dot{x}_2} \left(\dot{x}_1 \tanh{\left(\frac{x_{12}}{2} \right)} 
-\dot{x}_3 \tanh{\left(\frac{x_{23}}{2} \right)} \right) \theta_2
\nonumber\\[2pt]
&&
\qquad \quad 
-\sqrt{\dot{x}_3} \left(\dot{x}_1 \tanh{\left(\frac{x_{13}}{2} \right)} 
+\dot{x}_2 \tanh{\left(\frac{x_{23}}{2} \right)} \right) \theta_3.
\eea
Note that, when passing from the Hamiltonian formalism to the Lagrangian picture, $Q_2$ 
acquires a cubic fermionic contribution, whereas $Q_3$ loses such a term. This is a consequence of 
the explicit relations between $\lambda_i$ and $(x_i,{\dot x}_i,\theta_i)$ 
(see Appendix).

Within the Lagrangian framework, algebraic resolvability of the fermionic constants of the motion (\ref{FIOM}) 
is readily established. Because the cubic term $\Omega=\theta_1 \theta_2 \theta_3$ is  
conserved over time by itself,  Eqs. (\ref{FIOM})  can be put into
the linear algebraic form $a_{ij} \theta_j=b_i$, where $b_i$ is a specific vector function, 
which can be easily read off from (\ref{FIOM}), and $a_{ij}$ is a matrix involving three rows
\bea
&&
a_{1i}=\left(\sqrt{{\dot x}_1},\sqrt{{\dot x}_2},\sqrt{{\dot x}_3} \right), 
\nonumber\\[2pt]
&&
a_{2i}=\left(\sqrt{{\dot x}_2 {\dot x}_3} \tanh{\left(\frac{x_{23}}{2}\right)},
-\sqrt{{\dot x}_1 {\dot x}_3} \tanh{\left(\frac{x_{13}}{2}\right)},\sqrt{{\dot x}_1 {\dot x}_2} 
\tanh{\left(\frac{x_{12}}{2}\right)} \right),
\nonumber\\[2pt]
&&
a_{3i}=\epsilon_{ijk} a_{1j} a_{2k}, \qquad \vec{a}_3=\vec{a}_1 \times \vec{a}_2,
\eea
$\epsilon_{ijk}$ being the totally antisymmetric symbol with $\epsilon_{123}=1$. 
Because the determinant of $a_{ij}$ is equal to the square of the area of a parallelogram formed by 
the vectors $\vec{a}_1$ and $\vec{a}_2$, the matrix $a_{ij}$ is invertible and hence the system (\ref{FIOM}) 
is algebraically resolvable: $\theta_i={\left(a^{-1} \right)}_{ij} b_j$. Taking the resulting expressions 
and computing $\theta_1 \theta_2 \theta_3$, one can then link the Grassmann--odd constant of the motion 
$\Omega$ to $(Q_1,Q_2,Q_3)$ and $(\mathcal{I}_1,\mathcal{I}_2,\mathcal{I}_3)$.
 
\vspace{0.5cm}

\noindent
{\it 3.3 The Lax pair}\\

It is interesting to inquire how the $\mathcal{N}=1$ supersymmetric generalization
of the Ruijsenaars--Schneider three--body model constructed above modifies the original 
bosonic Lax pair (\ref{LP}). A straightforward way to find the modification is to 
add to each component of $L_{ij}$ and $M_{ij}$ in (\ref{LP}) extra terms quadratic 
in the fermionic degrees of freedom and then fix their explicit form by imposing 
identities analogous to (\ref{Lid})
\be\label{Lid1}
\mbox{Tr} L-\mathcal{I}_1=0, \qquad 
\mbox{Tr} L^2-{\left(\mbox{Tr} L\right)}^2+2 \mathcal{I}_2=0,
\qquad
\mbox{Tr} L^3-{\left(\mbox{Tr} L \right)}^3
+3 \mathcal{I}_1 \mathcal{I}_2-3 \mathcal{I}_3=0,
\ee
with $(\mathcal{I}_1,\mathcal{I}_2,\mathcal{I}_3)$ given in (\ref{LaFI}). 
For $L_{ij}$ the result reads
\bea
&&
L_{11}={\dot x}_1, \qquad L_{22}={\dot x}_2, \qquad L_{33}={\dot x}_3,
\nonumber\\[2pt]
&&
L_{12}=L_{21}=\frac{\sqrt{ {\dot x}_1 {\dot x}_2}}{\cosh{\left(\frac{x_{12}}{2} \right)}}
+\frac{{\rm i}}{8} \frac{\tanh{\left(\frac{x_{12}}{2} \right)}}{\cosh{\left(\frac{x_{12}}{2} \right)}}
\left( {\dot x}_1+ {\dot x}_2 \right) \theta_1 \theta_2
\nonumber\\[2pt]
&&
\qquad
+
\frac{{\rm i}}{8} 
\frac{\tanh{\left(\frac{x_{12}}{2} \right)}}
{\cosh{\left(\frac{x_{13}}{2} \right)}\cosh{\left(\frac{x_{23}}{2} \right)}}
\left(\sqrt{{\dot x}_2} \theta_1-\sqrt{{\dot x}_1} \theta_2 \right) \sqrt{{\dot x}_3} 
\theta_3,
\nonumber
\eea
\bea
&&
L_{13}=L_{31}=\frac{\sqrt{ {\dot x}_1 {\dot x}_3}}{\cosh{\left(\frac{x_{13}}{2} \right)}}
+\frac{{\rm i}}{8} \frac{\tanh{\left(\frac{x_{13}}{2} \right)}}
{\cosh{\left(\frac{x_{13}}{2} \right)}}
\left( {\dot x}_1+ {\dot x}_3 \right) \theta_1 \theta_3
\nonumber\\[2pt]
&&
\qquad
+
\frac{{\rm i}}{8} 
\frac{\tanh{\left(\frac{x_{13}}{2} \right)}}
{\cosh{\left(\frac{x_{12}}{2} \right)}\cosh{\left(\frac{x_{23}}{2} \right)}}
\left(\sqrt{{\dot x}_3} \theta_1-\sqrt{{\dot x}_1} \theta_3 \right) 
\sqrt{{\dot x}_2} \theta_2,
\nonumber\\[2pt]
&&
L_{23}=L_{32}=\frac{\sqrt{ {\dot x}_2 {\dot x}_3}}{\cosh{\left(\frac{x_{23}}{2} \right)}}
+\frac{{\rm i}}{8} \frac{\tanh{\left(\frac{x_{23}}{2} \right)}}
{\cosh{\left(\frac{x_{23}}{2} \right)}}
\left( {\dot x}_2+ {\dot x}_3 \right) \theta_2 \theta_3
\nonumber\\[2pt]
&&
\qquad
+
\frac{{\rm i}}{8} 
\frac{\tanh{\left(\frac{x_{23}}{2} \right)}}
{\cosh{\left(\frac{x_{12}}{2} \right)}\cosh{\left(\frac{x_{13}}{2} \right)}}
\left(\sqrt{{\dot x}_3} \theta_2-\sqrt{{\dot x}_2} \theta_3 \right) 
\sqrt{{\dot x}_1} \theta_1,
\eea
while the superextension of $M_{ij}$ in (\ref{LP}) takes the form ($M_{11}=M_{22}=M_{33}=0$)
\bea
&&
M_{12}=-M_{21}=-\frac{\sqrt{ {\dot x}_1 {\dot x}_2}}{2\sinh{\left(\frac{x_{12}}{2} \right)}}
-\frac{{\rm i} \left( {\dot x}_1+ {\dot x}_2 \right) }{16\cosh{\left(\frac{x_{12}}{2} \right)}}
 \theta_1 \theta_2
\nonumber\\[2pt]
&&
\qquad
- 
\frac{{\rm i}}
{16 \cosh{\left(\frac{x_{13}}{2} \right)}\cosh{\left(\frac{x_{23}}{2} \right)}}
\left(\sqrt{{\dot x}_2} \theta_1-\sqrt{{\dot x}_1} \theta_2 \right) \sqrt{{\dot x}_3} \theta_3,
\nonumber\\[2pt]
&&
M_{13}=-M_{31}=-\frac{\sqrt{ {\dot x}_1 {\dot x}_3}}{2\sinh{\left(\frac{x_{13}}{2} \right)}}
-\frac{{\rm i} \left( {\dot x}_1+ {\dot x}_3 \right) }{16 \cosh{\left(\frac{x_{13}}{2} \right)}}
 \theta_1 \theta_3
\nonumber\\[2pt]
&&
\qquad
-
\frac{{\rm i}}
{16\cosh{\left(\frac{x_{12}}{2} \right)}\cosh{\left(\frac{x_{23}}{2} \right)}}
\left(\sqrt{{\dot x}_3} \theta_1-\sqrt{{\dot x}_1} \theta_3 \right) 
\sqrt{{\dot x}_2} \theta_2,
\nonumber\\[2pt]
&&
M_{23}=-M_{32}=-\frac{\sqrt{ {\dot x}_2 {\dot x}_3}}{2\sinh{\left(\frac{x_{23}}{2} \right)}}
-\frac{{\rm i} \left( {\dot x}_2+ {\dot x}_3 \right) }{16 \cosh{\left(\frac{x_{23}}{2} \right)}}
 \theta_2 \theta_3
\nonumber\\[2pt]
&&
\qquad
-
\frac{{\rm i}}
{16\cosh{\left(\frac{x_{12}}{2} \right)}\cosh{\left(\frac{x_{13}}{2} \right)}}
\left(\sqrt{{\dot x}_3} \theta_2-\sqrt{{\dot x}_2} \theta_3 \right) 
\sqrt{{\dot x}_1} \theta_1.
\eea
It is straightforward to verify that the matrix equation ${\dot L}=[L,M]$ is satisfied, 
provided the  
supersymmetric equations of motion (\ref{SUSYeq1}) and  (\ref{SUSYeq2}) hold.

The issue of constructing a Lax pair in the fermionic sector appears to be more subtle. 
As was explained above, only those fermionic integrals of motion matter, 
which involve contributions linear in the fermionic degrees of freedom.  
If ${\dot R}=[R,S]$, where $R_{ij}$ is a Grassmann--odd matrix and $S_{ij}$ 
is a Grassmann--even matrix, is going to represent a Lax pair in the fermionic sector,
$\mbox{Tr} R$ should link to one of integrals of motion relevant for the algebraic resolvability 
in the fermionic sector, while $\mbox{Tr} R^k$, 
$k=2,3,\dots$, would play no essential role. 
Given the fact that one has three fermionic constants of the motion displayed in Eqs. (\ref{FIOM}) above,
one expects to find more than one Lax pair in the fermionic sector. 
Postponing a detailed analysis of this issue for future study, below we display a variant which links to $Q_1$ 
in (\ref{FIOM}) ($S_{11}=S_{22}=S_{33}=0$)
\bea
&&
R_{11}=\sqrt{{\dot x}_1} \theta_1, \qquad R_{22}=\sqrt{{\dot x}_2} \theta_2, \qquad 
R_{33}=\sqrt{{\dot x}_3} \theta_3, 
\nonumber
\eea
\bea
&&
R_{12}=R_{21}=\frac 12 \left(\sqrt{{\dot x}_2} \theta_1 +\sqrt{{\dot x}_1} \theta_2\right),
\nonumber\\[2pt]
&&
R_{13}=R_{31}=\frac 12 \left(\sqrt{{\dot x}_3} \theta_1 
+\sqrt{{\dot x}_1} \theta_3\right), 
\nonumber\\[2pt]
&&
R_{23}=R_{32}=\frac 12 \left(\sqrt{{\dot x}_3} \theta_2 
+\sqrt{{\dot x}_2} \theta_3\right),
\nonumber\\[2pt]
&&
S_{12}=-S_{21}=-\frac{\sqrt{{\dot x}_1 {\dot x}_2}}{\sinh{x_{12}}}
-\frac{{\rm i} \tanh{\left(\frac{x_{13}}{2} \right)} 
\tanh{\left(\frac{x_{23}}{2} \right)} \left({\dot x}_1 +{\dot x}_2 \right)}
{8 \cosh^2 {\left(\frac{x_{12}}{2} \right)}} \theta_1 \theta_2, 
\nonumber\\[2pt]
&&
S_{13}=-S_{31}=-\frac{\sqrt{{\dot x}_1 {\dot x}_3}}{\sinh{x_{13}}}
+\frac{{\rm i} \tanh{\left(\frac{x_{12}}{2} \right)} 
\tanh{\left(\frac{x_{23}}{2} \right)} \left({\dot x}_1 +{\dot x}_3 \right)}
{8 \cosh^2 {\left(\frac{x_{13}}{2} \right)}} \theta_1 \theta_3, 
\nonumber\\[2pt]
&&
S_{23}=-S_{32}=-\frac{\sqrt{{\dot x}_2 {\dot x}_3}}{\sinh{x_{23}}}
-\frac{{\rm i} \tanh{\left(\frac{x_{12}}{2} \right)} 
\tanh{\left(\frac{x_{13}}{2} \right)} \left({\dot x}_2 +{\dot x}_3 \right)}
{8 \cosh^2 {\left(\frac{x_{23}}{2} \right)}} \theta_2 \theta_3.
\eea
One can readily verify that the equation ${\dot R}=[R,S]$ holds, provided (\ref{SUSYeq1}) 
and  (\ref{SUSYeq2}) are satisfied and $\mbox{Tr} R=Q_1$.

\vspace{0.5cm}

\noindent
{\bf 4. Integrable isospin extension of the Ruijsenaars--Schneider three--body model}\\

At the classical level, bosonic and fermionic degrees of freedom, which characterize 
a supersymmetric mechanics, take values in the infinite--dimensional Grassmann algebra \cite{BK}. 
For this reason, their mechanical interpretation is somewhat obscure. 
As was advocated in a recent work \cite{AG1}, a supersymmetrization 
of a given integrable system can be used to construct an integrable extension of 
the original model by (bosonic) isospin degrees of freedom. In this section, we apply 
the procedure in \cite{AG1} to the Ruijsenaars--Schneider three--body model.

According to our analysis above, the fermionic sector of the $\mathcal{N}=1$ 
supersymmetric extension of the Ruijsenaars--Schneider three--body model 
is described by three Grassmann--odd variables
$\theta_i$, $i=1,2,3$, which obey the first order differential equations (\ref{SUSYeq2}).
The corresponding general solution involves three Grassmann--odd constants of integration.
Denoting them by $\alpha$, $\beta$, and $\gamma$ and taking into account the equalities
$\alpha^2=\beta^2=\gamma^2=0$, one arrives at the following natural decompositions
\bea\label{comp}
&&
\theta_i=\alpha \varphi_{i1} + \beta \varphi_{i2}+\gamma \varphi_{i3} + 
{\rm i} \alpha \beta \gamma \varphi_{i4}, \qquad
x_i =x_{i 0} + {\rm i} \alpha\beta x_{ i 1}  +{\rm i} \alpha\gamma x_{ i 2}  
+{\rm i} \beta\gamma x_{ i 3},
\eea
where components accompanying $\alpha$, $\beta$, and $\gamma$ 
are real {\it bosonic} functions of the temporal variable $t$. 
Substituting (\ref{comp}) into the equations of motion (\ref{SUSYeq1}), (\ref{SUSYeq2})
and analyzing monomials in $\alpha$, $\beta$, $\gamma$ on both sides,
one can turn (\ref{SUSYeq1}), (\ref{SUSYeq2}) into a system of ordinary 
differential equations for usual real--valued functions. The latter provides 
an integrable extension of the original bosonic model.

In general, the resulting system is rather bulky and hard to interpret. 
Yet, a simple and tractable
extension arises 
if one focuses on a particular solution for which $\beta=\gamma=0$
\bea\label{comp1}
&&
\theta_i=\alpha \varphi_{i}, \qquad \alpha^2=0,
\eea
$\varphi_{i}$ being a real--valued bosonic function to be interpreted below as describing 
isospin degrees of freedom.
In this case, all terms quadratic (or higher) in the fermionic variables $\theta_i$ 
vanish and the $\mathcal{N}=1$ superextension reduces to the original 
Ruijsenaars--Schneider equations (\ref{RS}) accompanied by the linear differential 
equations
\be\label{ISO}
\dot{\varphi}_i=\frac 12 \sum_{j\ne i} W(x_{ij}) \sqrt{{\dot x}_i {\dot x}_j}  
 {\varphi}_j,
\ee
which govern the evolution of the isospin degrees of freedom $\varphi_i$. 
Given the structure of the equations (\ref{ISO}), one can readily establish that
\be
I_4=\varphi_i \varphi_i
\ee
is conserved over time. Thus, the internal degrees of 
freedom $\varphi_i$ parameterize a two--sphere. 

Taking into account the explicit form of $(Q_1,Q_2,Q_3)$ in (\ref{FIOM}) and the assumption 
made in (\ref{comp1}), 
one then finds three first integrals of (\ref{ISO}) for $W(x)=\frac{2}{\sinh{x}}$
\bea\label{Int}
&&
I_5=\sqrt{{\dot x}_1} \varphi_1+\sqrt{{\dot x}_2} \varphi_2+\sqrt{{\dot x}_3} \varphi_3,
\nonumber\\[2pt]
&&
I_6=\sqrt{{\dot x}_2 {\dot x}_3} \tanh{\left(\frac{x_{23}}{2}\right)} \varphi_1
-\sqrt{{\dot x}_1 {\dot x}_3} \tanh{\left(\frac{x_{13}}{2}\right)} \varphi_2
+\sqrt{{\dot x}_1 {\dot x}_2} \tanh{\left(\frac{x_{12}}{2}\right)} \varphi_3,
\nonumber\\[2pt]
&&
I_7=\sqrt{\dot{x}_1} \left(\dot{x}_2 \tanh{\left(\frac{x_{12}}{2} \right)} 
+\dot{x}_3 \tanh{\left(\frac{x_{13}}{2} \right)} \right) \varphi_1
\nonumber\\[2pt]
&&
\qquad \quad 
-\sqrt{\dot{x}_2} \left(\dot{x}_1 \tanh{\left(\frac{x_{12}}{2} \right)} 
-\dot{x}_3 \tanh{\left(\frac{x_{23}}{2} \right)} \right) \varphi_2
\nonumber\\[2pt]
&&
\qquad \quad 
-\sqrt{\dot{x}_3} \left(\dot{x}_1 \tanh{\left(\frac{x_{13}}{2} \right)} 
+\dot{x}_2 \tanh{\left(\frac{x_{23}}{2} \right)} \right) \varphi_3.
\eea
$(I_4,I_5,I_6,I_7)$ given above along with $(I_1,I_2,I_3)$ in (\ref{fi}) provide seven 
functionally independent integrals of motion characterising the isospin extension of the 
Ruijsenaars--Schneider three--body model (\ref{RS}), (\ref{ISO}). 
More complicated examples can be 
constructed by keeping $\beta$, or $\gamma$, or both nonzero.

\vspace{0.5cm}

\noindent
{\bf 5. Conclusion}\\

To summarize, in this work integrability of an $\mathcal{N}=1$ 
supersymmetric Ruijsenaars-Schneider three--body model was established. 
Three functionally independent Grassmann--odd constants of the motion 
were constructed and their algebraic resolvability was proven.
The supersymmetric generalization was used to build a novel 
integrable isospin extension of the Ruijsenaars-Schneider three--body system.

Turning to issues deserving of further investigation, an interesting open problem is
to extend the present analysis to the case of more than three interacting particles.
A preliminary consideration shows that it might not be that easy. It is straightforward 
to construct  four subsidiary functions $\lambda_i$ similar to those in (\ref{sf})
\bea\label{SF}
&&
\lambda_1=e^{\frac{p_1}{2}} \sqrt{ \coth{\left(\frac{x_{12}}{2}\right)}
\coth{\left(\frac{x_{13}}{2}\right)} \coth{\left(\frac{x_{14}}{2}\right)}}, 
\nonumber\\[2pt]
&& 
\lambda_2=e^{\frac{p_2}{2}}\sqrt{ \coth{\left(\frac{x_{12}}{2}\right)} 
\coth{\left(\frac{x_{23}}{2}\right)} \coth{\left(\frac{x_{24}}{2}\right)}}, 
\nonumber\\[2pt]
&&
\lambda_3=e^{\frac{p_3}{2} } \sqrt{\coth{\left(\frac{x_{13}}{2}\right)}
\coth{\left(\frac{x_{23}}{2}\right)} \coth{\left(\frac{x_{34}}{2}\right)}}, 
\nonumber\\[2pt]
&&
\lambda_4=e^{\frac{p_4}{2} } \sqrt{\coth{\left(\frac{x_{14}}{2}\right)}
\coth{\left(\frac{x_{24}}{2}\right)} \coth{\left(\frac{x_{34}}{2}\right)}},
\eea
which obey the quadratic algebra (\ref{All}) for the four--body case.
An $\mathcal{N}=1$ supersymmetry charge can be chosen as before: 
$Q_1=\lambda_i \theta_i$, $i=1,\dots,4$. In order to ensure integrability in the 
fermionic sector parameterized by $\theta_i$, one has to construct three more 
Grassmann--odd constants of the motion, which form an algebraically resolvable 
set jointly with $Q_1$. Note that in the three--body case the structure of
the second fermionic invariant $Q_2$ in (\ref{QQ2}) was suggested by the fact that 
the original bosonic first integral $I_2$ in (\ref{fi}), to which $Q_2$ is the 
superpartner,
involved exactly three terms. Their number matches the number of fermions at hand. Likewise, for the four--body case, $I_3$ in (\ref{fi1}) contains 
four terms, which allow one to find the superpartner
\bea
&&
Q_3=\lambda_2 \lambda_3 \lambda_4 \tanh{\left(\frac{x_{23}}{2}\right)}  
\tanh{\left(\frac{x_{24}}{2}\right)} \tanh{\left(\frac{x_{34}}{2}\right)} \theta_1
\nonumber\\[2pt]
&&
\qquad \quad -
\lambda_1 \lambda_3 \lambda_4 \tanh{\left(\frac{x_{13}}{2}\right)}  
\tanh{\left(\frac{x_{14}}{2}\right)} \tanh{\left(\frac{x_{34}}{2}\right)} \theta_2
\nonumber\\[2pt]
&&
\qquad \quad
+\lambda_1 \lambda_2 \lambda_4 \tanh{\left(\frac{x_{12}}{2}\right)}  
\tanh{\left(\frac{x_{14}}{2}\right)} \tanh{\left(\frac{x_{24}}{2}\right)} \theta_3
\nonumber\\[2pt]
&&
\qquad \quad
-\lambda_1 \lambda_2 \lambda_3 \tanh{\left(\frac{x_{12}}{2}\right)}  
\tanh{\left(\frac{x_{13}}{2}\right)} \tanh{\left(\frac{x_{23}}{2}\right)} \theta_4.
\eea
In order to construct two more fermionic first integrals needed for integrability in the fermionic sector, 
one could proceed like in Sect. 3.1. A direct inspection of fermionic equations of motion shows that the 
quartic combination
\be
\Omega_4=\theta_1 \theta_2 \theta_3 \theta_4 
\ee
is conserved over time. Successive bracketing of $Q_1$ and $Q_3$ with $\Omega_4$ 
should result in new constants of the motion. Surprisingly enough, the explicit calculation 
yields extra cubic and quadratic descendents but no linear ones! 
Thus an alternative approach is yet to be developed.

Another issue deserving of further study is to understand whether each Grassmann--odd 
integral 
of motion entering (\ref{FIOM}) admits a Lax representation. Also worth investigating is how the 
Lax pairs in the bosonic and fermionic sectors transform under the $\mathcal{N}=1$ supersymmetry 
transformations.

A generalization of the present study to encompass various supersymmetric extensions 
of the Calogero model is an interesting avenue to explore.

In this work, we elaborated on the hyperbolic potential
$W(x)=\frac{2}{\sinh{x}}$. The case $W(x)=2 \coth{x}$ is worth studying as well. This last issue appears to be
purely technical however.

\vspace{0.5cm}

\noindent{\bf Acknowledgements}\\

\noindent
This work was supported by the Russian Science Foundation, grant No 23-11-00002.
\vspace{0.5cm}

\noindent
{\bf Appendix. Equations of motion for the subsidiary functions $\lambda_i$}\\

In this Appendix, we display the Hamiltonian equations of motion for the canonical coordinates $x_i$ and the subsidiary 
functions $\lambda_i$, which are extensively used for obtaining the Lagrangian equations of motion in Sect. 2
\bea\label{eqx}
&&
{\dot x}_1={\left(\lambda_1+\frac{{\rm i}}{4 \sinh{x_{12}}}   \lambda _2 \theta_1 \theta_2
+\frac{{\rm i}}{4 \sinh{x_{13}}}  \lambda _3 \theta_1 \theta_3 \right)}^2,
\nonumber\\[2pt]
&&
{\dot x}_2={\left(\lambda_2+\frac{{\rm i}}{4 \sinh{x_{12}}}  \lambda _1 \theta_1 \theta_2
+\frac{{\rm i}}{4 \sinh{x_{23}}}  \lambda _3 \theta_2 \theta_3 \right)}^2,
\nonumber\\[2pt]
&&
{\dot x}_3={\left(\lambda_3+\frac{{\rm i}}{4 \sinh{x_{13}}}  \lambda _1 \theta_1 \theta_3
+\frac{{\rm i}}{4 \sinh{x_{23}}}  \lambda _2 \theta_2 \theta_3 \right)}^2,
\nonumber\\[2pt]
&&
{\dot\lambda}_1=\frac{1}{\sinh{x_{12}}} \lambda_1 \lambda_2^2 +\frac{1}{\sinh{x_{13}}} \lambda_1 \lambda_3^2 +
\frac{{\rm i}}{2} \lambda_2 \lambda_1^2 \left(\frac{1+\cosh{x_{12}}}{\sinh^2{x_{12}}} \right) \theta_1 \theta_2
\nonumber\\[2pt]
&&
\qquad 
+
\frac{{\rm i}}{2} \lambda_3 \lambda_1^2 \left(\frac{1+\cosh{x_{13}}}{\sinh^2{x_{13}}} \right) \theta_1 \theta_3
+\frac{{\rm i}}{2} \lambda_1 \lambda_2 \lambda_3 \frac{1}{\sinh{x_{23}}} \left(\frac{1}{\sinh{x_{12}}}
+\frac{1}{\sinh{x_{13}}}\right) \theta_2 \theta_3,
\nonumber\\[2pt]
&&
{\dot\lambda}_2=-\frac{1}{\sinh{x_{12}}} \lambda_2 \lambda_1^2 +\frac{1}{\sinh{x_{23}}} \lambda_2 \lambda_3^2 -
\frac{{\rm i}}{2} \lambda_1 \lambda_2^2 \left(\frac{1+\cosh{x_{12}}}{\sinh^2{x_{12}}} \right) \theta_1 \theta_2
\nonumber\\[2pt]
&&
\qquad 
+\frac{{\rm i}}{2} \lambda_3 \lambda_2^2 \left(\frac{1+\cosh{x_{23}}}{\sinh^2{x_{23}}} \right) \theta_2 \theta_3
-\frac{{\rm i}}{2} \lambda_1 \lambda_2 \lambda_3 \frac{1}{\sinh{x_{13}}}
\left(\frac{1}{\sinh{x_{12}}}-\frac{1}{\sinh{x_{23}}}\right) \theta_1 \theta_3,
\nonumber\\[2pt]
&&
{\dot\lambda}_3=-\frac{1}{\sinh{x_{13}}} \lambda_3 \lambda_1^2 -\frac{1}{\sinh{x_{23}}} \lambda_3 \lambda_2^2 -
\frac{{\rm i}}{2} \lambda_1 \lambda_3^2 \left(\frac{1+\cosh{x_{13}}}{\sinh^2{x_{13}}} \right) \theta_1 \theta_3
\nonumber\\[2pt]
&&
\qquad 
-\frac{{\rm i}}{2} \lambda_2 \lambda_3^2 \left(\frac{1+\cosh{x_{23}}}{\sinh^2{x_{23}}} \right) \theta_2 \theta_3
-\frac{{\rm i}}{2} \lambda_1 \lambda_2 \lambda_3 \frac{1}{\sinh{x_{12}}}
\left(\frac{1}{\sinh{x_{13}}}+\frac{1}{\sinh{x_{23}}}\right) \theta_1 \theta_2.
\nonumber
\eea

\end{document}